# Bias Tunable Transport Modulation and Gas Selectivity in Layered BiOI: A DFT NEGF Study

Jemal Yimer Damte [a], Jiří Houška [a], Pavel Baroch [a], Xue Yong [b]

[a] Department of Physics and NTIS – European Centre of Excellence, University of West Bohemia in Pilsen, Univerzitni 8 30100 Plzen, Czech Republic

[b] Department of Electrical Engineering and Electronics, University of Liverpool, Liverpool L69 3GJ, UK

## Abstract

Understanding the interplay between adsorption energetics and charge transport modulation is essential for the rational design of low-power and bias-tunable gas sensors. Here, we present a comprehensive first principles study of gas selectivity in layered bismuth oxyiodide (BiOI) by integrating density functional theory with non-equilibrium Green's function transport calculations. The adsorption and bias dependent transport responses toward $NO_2$, $NH_3$, $CO_2$, and representative volatile organic compounds are systematically examined. While $NH_3$ and $NO_2$ exhibit strong chemisorption and localized electronic perturbation, $CO_2$ interacts via weak physisorption, demonstrating that adsorption strength alone does not determine sensing performance. Instead, transmission channel evolution near the Fermi level governs the sensing response. Bias dependent calculations reveal an electrically tunable sensitivity hierarchy, where weakly adsorbed $CO_2$ preserves conductive pathways and exhibits pronounced low bias sensitivity despite minimal charge transfer. Recovery time analysis further highlights the trade-off between transport modulation and reversibility for strongly adsorbed species. These results establish a transport centered selectivity framework for layered BiOI and provide mechanistic insight into electric field-controlled gas sensing under ambient conditions.

# 1. Introduction

The detection of hazardous gases such as nitrogen dioxide ($NO_2$), ammonia ($NH_3$), carbon dioxide ($CO_2$), and volatile organic compounds (VOCs) is essential for environmental monitoring, industrial safety, and public health protection[1–4] These gases originate from combustion processes, agricultural activities, chemical industries, and indoor pollution sources, and are associated with respiratory disorders, atmospheric chemistry imbalance, and climate change[5–8]. Reliable detection under ambient conditions is therefore a critical requirement for next generation sensing technologies. Conventional chemiresistive gas sensors based on metal oxide semiconductors typically rely on thermally activated surface reactions to generate measurable electrical signals[9–11]. Elevated operating temperatures (200–400 °C) are commonly required to promote adsorption desorption kinetics and redox processes, leading to high power consumption, baseline drift, and limited integration into portable or wearable platforms[12–16]. Consequently, significant research efforts have been devoted to the development of low power gas sensors that operate through electronic modulation rather than thermally driven chemical reactions[17–19].

Layered semiconductors have emerged as promising candidates for gas sensing due to their high surface to volume ratio, exposed active sites, and tunable electronic structures[20–22]. Among these materials, bismuth oxyiodide (BiOI), a member of the BiOX (X = Cl, Br, I) family, possesses a unique layered structure composed of alternating $[Bi_2O_2]^{2+}$ slabs and iodide layers. This architecture generates an intrinsic internal electric field perpendicular to the layers, facilitating charge separation and surface polarization[23–25]. In addition, BiOI exhibits p-type semiconducting behavior and a relatively narrow band gap, both of which favor adsorption induced charge transfer and conductivity modulation[21,26,27]. Although BiOI has been extensively investigated for photocatalytic and optoelectronic applications, its gas sensing properties remain comparatively underexplored. Existing studies often focus on single target gases or emphasize adsorption energetics as the primary descriptor of sensing performance[28,29]. However, adsorption strength alone does not necessarily correlate with electrical sensitivity. Strong chemisorption may enhance charge transfer but can simultaneously introduce localized states that suppress carrier transport and hinder recovery[30–33]. Conversely, weak physisorption may preserve conductive pathways while still enabling measurable signal modulation. Therefore, a comprehensive understanding of gas sensing in layered semiconductors requires coupling adsorption physics with charge transport analysis under applied bias[34,35]. Ab initio calculations provide atomistic insight into the sensing mechanisms of various gas sensors, as demonstrated in numerous previous studies[36,37].

Previous experimental studies have demonstrated the gas-sensing potential of BiOI and BiOI-containing materials. Sun and Amrillah reviewed the structural, electronic, and optoelectronic properties of BiOI, highlighting its potential for light-assisted sensing and related applications [38]. Navale et al. investigated mesoporous BiOI nanoflakes for chemiresistive ethanol sensing[35], while Li et al. demonstrated room-temperature $NO_2$ detection using BiOI-functionalized ZnO nanorods under visible-light illumination[34]. Theoretical investigations have also established the usefulness of combining density functional theory (DFT) with nonequilibrium Green's function (NEGF) calculations to examine gas-induced changes in electronic transport. For example, Chen and Yeh investigated the adsorption and sensing behavior of nitrogen-containing gases on a $B_2CO$ monolayer using DFT–NEGF calculations[39]. More recently, Chen et al. combined experimental characterization with DFT calculations to investigate $NO_2$ sensing in defect-engineered BiOI/$MoS_2$ heterostructures[40]. Beyond individual sensing materials, advanced gas-detection systems increasingly combine sensor arrays with computational approaches to improve gas identification. Wang et al. developed a portable electronic-nose system integrating edge computing, ensemble learning, and a sensor-swarm architecture[41]. Zhu et al. demonstrated multimodal intelligent MEMS gas sensors based on NiO/ZnO nanocomposites, incorporating deep learning for VOC identification and concentration prediction[42]. These developments highlight the importance of distinguishing gas-specific electrical responses, providing broader motivation for investigating bias-dependent transport fingerprints in BiOI.

Although several experimental studies[34,35] have demonstrated the potential of BiOI and BiOI-containing materials for gas sensing, the literature remains relatively limited, with most studies focusing on specific target gases and heterostructure or defect engineering. Furthermore, the aforementioned literature is predominantly experimental, with almost no related theoretical support. Thus, we present a systematic first principles investigation of gas sensing mechanisms within a BiOI(001) DFT–NEGF device configuration by combining density functional theory (DFT) with non-equilibrium Green's function (NEGF) transport calculations. The adsorption behavior, electronic structure modification, and transport responses toward $NO_2$, $NH_3$, $CO_2$, and representative VOCs including $CH_3OH$, $CH_3COCH_3$, and $C_2H_5OH$ are comparatively analyzed as a function of applied source–drain bias. Rather than ranking gases solely by adsorption energy, we demonstrate that the sensing hierarchy can be electrically tuned through bias controlled modulation of transmission channels. Notably, weakly adsorbed $CO_2$ preserves conductive pathways near the Fermi level and exhibits pronounced low bias sensitivity, demonstrating that bias dependent transport modulation rather than adsorption strength alone governs gas selectivity in layered BiOI. By establishing a correlation between adsorption induced charge redistribution, density of states perturbation, and field assisted carrier transport, this study provides a transport centered framework for multi gas selectivity in layered semiconductors. The insights gained here offer fundamental guidance for the rational design of low power, bias tunable gas sensors operating under ambient conditions.

## 2. Computational Details

First principles calculations were carried out within the framework of density functional theory (DFT) using the Vienna *Ab Initio* Simulation Package (VASP) [43]. The electronic exchange correlation interactions were treated using the generalized gradient approximation (GGA) with the Perdew–Burke–Ernzerhof (PBE) functional[44]. Core valence electron interactions were described by the projector augmented wave (PAW) method[45]. To model the BiOI system, a 3x3 supercell was constructed, which is sufficiently large to accommodate surface adsorption and defect induced structural relaxations. The plane wave basis set was truncated at a kinetic energy cutoff of 450 eV. Sampling of the Brillouin zone was performed using a Monkhorst–Pack k-point grid of $6 \times 6 \times 1$ and spin polarization have been considered. To avoid spurious interactions between periodically repeated slabs, a vacuum layer of 15 Å was introduced along the surface normal direction. Structural optimization was conducted by fully relaxing all atomic coordinates until the residual forces on each atom were less than 0.01 eV Å$^{-1}$, while the total energy convergence threshold was set to $1 \times 10^{-6}$ eV.

Convergence tests were performed for bulk BiOI to assess the k-point sampling and plane-wave energy cutoff. The total energy was −26.47 eV for the $6 \times 6 \times 6$, $7 \times 7 \times 7$, and $8 \times 8 \times 8$ k-point meshes at the reported numerical precision (Table S1). Increasing the plane-wave cutoff from 450 to 500 eV changed the total energy by 0.01 eV per simulation cell. At 450 eV, the calculated lattice parameters (a = b = 3.99 Å and c = 9.12 Å) agree closely with the experimental values (a = b = 3.99 Å and c = 9.21 Å) reported by Jagt et al. [46] and are comparable to their calculated static-lattice values (a = b = 3.97 Å and c = 9.17 Å). Accordingly, 450 eV was retained for the comparative calculations, while the remaining cutoff dependence of the out-of-plane lattice parameter is acknowledged. Furthermore, to assess the influence of the exchange–correlation functional, the spin-resolved density of states of pristine BiOI(001) was calculated using both PBE and HSE06 (Figure S2b). While the presented PBE-based values and especially trends arguably support the presented conclusions (mechanism allowing the sensors to be selective due to different dependencies of sensor sensitivity on the bias voltage for different molecules), the comparison in Figure. S2b nevertheless reveals differences in the positions of individual electronic states, demonstrating the sensitivity of the quantitative information to the functional employed.

The BiOI(001) slab was constructed from layered I–Bi–O–Bi–I structural units separated by an internal I···I van der Waals gap (Figure 1). The exposed surface was modified to obtain the O-rich termination used in the adsorption study. No mirror-symmetry constraint was imposed during the structural relaxation.

To account for long range dispersion interactions that are not adequately described by conventional GGA functionals, van der Waals corrections were incorporated using the DFT-D3 method[47]. Adsorption energetics were quantified by calculating the adsorption energy ($E_{ads}$), defined as

$$E_{ads} = E_{Surface+molecule} - E_{Surface} - E_{molecule}$$

where $E_{Surface+molecule}$ is the total energy of the surface with an adsorbed gas molecule, $E_{Surface}$ corresponds to the clean surface energy, and $E_{molecule}$ represents the total energy of the isolated gas molecule. The electronic transport properties of the sensing systems were analyzed by calculating the I–V characteristics with the TranSIESTA code within the framework of the NEGF method[48]. In the transport setup, gold served as the electrode material, while the device region included a scattering zone connected to semi-infinite electrodes on both sides. The resulting transport current was determined using the Landauer–Büttiker formalism[49]. The combination of DFT and NEGF provides a robust theoretical framework for gas sensing studies. This integrated approach allows for a direct correlation between gas adsorption phenomena and transport responses, which is critical for understanding and optimizing the sensitivity and selectivity of gas-sensing materials.

Transport was calculated self-consistently using TranSIESTA in a two-probe Au–BiOI–Au geometry. The central scattering region contains the BiOI sensing layer and six explicit Au atomic planes at each contact, connected to semi-infinite Au electrodes along the $\pm a_3$ directions. The supplied device input uses a DZP basis, a 700 eV real-space mesh cutoff, a $5 \times 1 \times 11$ k-point grid, and an electronic temperature of 50 meV. The current is obtained by integrating the bias-dependent transmission $T(E, V)$ between the electrode electrochemical potentials, with transmission post-processing performed using TBTrans. The current was calculated using the Landauer–Büttiker formalism[49], in which the current is obtained by integrating the energy-dependent transmission function:

$$\frac{2e}{\hbar}\int_{-\infty}^{\infty} T(E,V_b)[f_L(E - \mu_L) - f_R(E - \mu_R)]\, dE = I(V_b)$$

Here, $f_L$ and $f_R$ denote the Fermi–Dirac distribution functions of the left and right electrodes, respectively; e is the elementary charge; $I(V_b)$ is the electric current at the applied bias voltage $V_b$, $\mu_L$ and $\mu_R$ are the chemical potentials of the left and right electrodes, respectively; and $\hbar$ is the reduced Planck constant. Under an applied bias, the chemical potentials of the electrodes are shifted relative to the Fermi energy, $E_F$, according to

$$\left(\mu_{L(R)} = E_F \pm \frac{e\, V_b}{2}\right)$$

The energy- and bias-dependent transmission coefficient, $T(E, V_b)$, was calculated as

$$T(E,V_b) = \mathrm{Tr}[\Gamma_L(E,V_b)G(E,V_b)\Gamma_R(E,V_b)G^*(E,V_b)]$$

In this expression, $G(E,V_b)\Gamma_R$ and $G^*(E,V_b)]$ represent the retarded and advanced Green's functions, respectively, while $\Gamma_L(E,V_b)$ and $\Gamma_R(E,V_b)G^*$ denote the contact-broadening functions associated with the left and right electrodes, describing the coupling between the transport region and the electrodes[50].

The pristine and gas-adsorbed BiOI geometries were relaxed using VASP and transferred to the TranSIESTA two-probe model without additional Au–BiOI interface relaxation. The central scattering region contains BiOI between left and right semi-infinite fcc Au electrodes with (001)-type contact planes; the transport axis is $a_3$, normal to these planes. Six explicit Au atomic planes are included on each side. For the constructed device geometry in the supplied transport input, the separations along $a_3$ between the terminal Au planes and nearest BiOI atomic planes are approximately 2.79 Å (left) and 2.85 Å (right); The full Au–BiOI–Au transmission includes electronic effects of both interfaces, but separate contact and channel resistances were not extracted. Consequently, the absolute current depends on the adopted interface geometry, and neither the absence of a contact barrier nor contact-independent resistance is inferred. Let us emphasize that the same electrode construction protocol was applied to pristine and gas-adsorbed devices when comparing relative current changes.

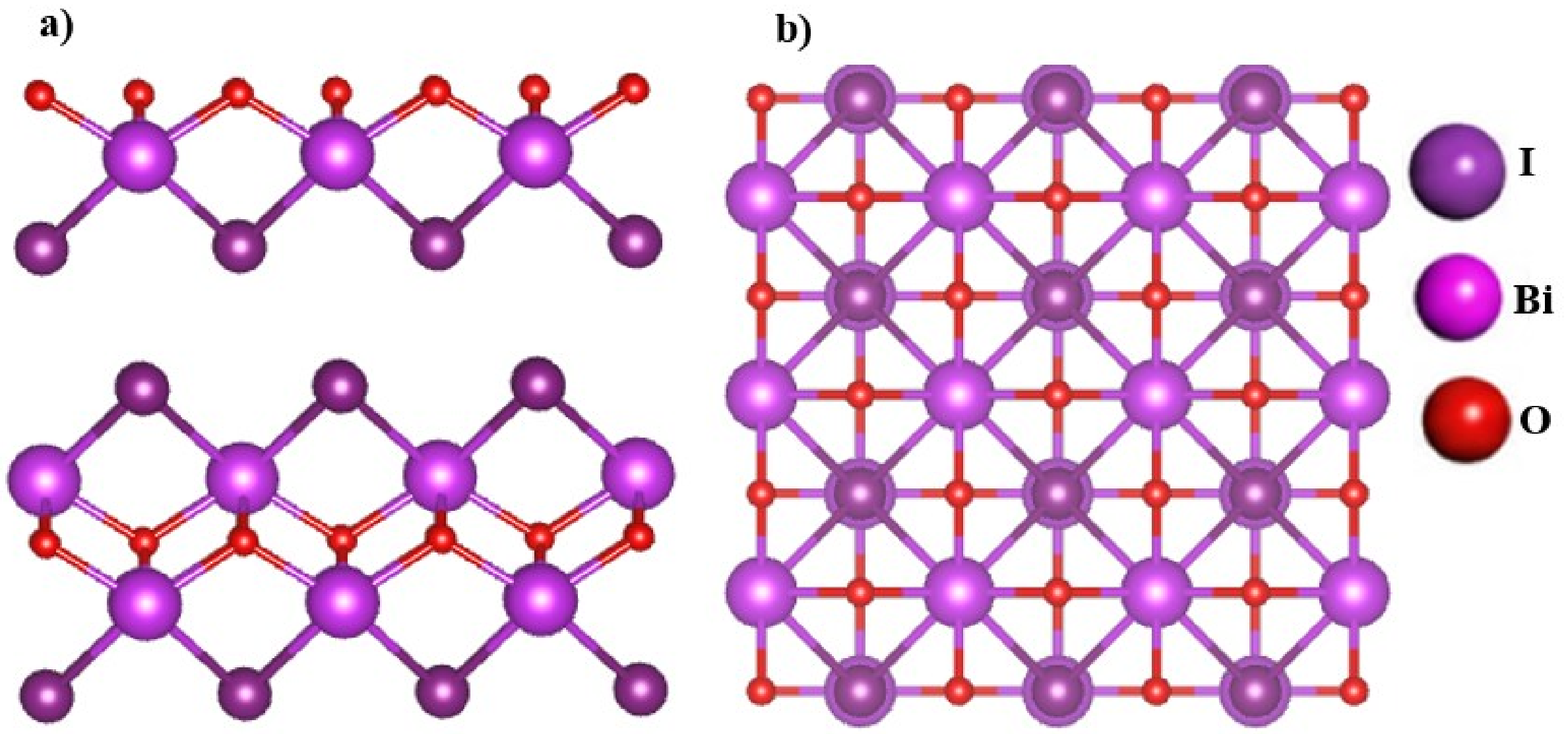

Figure 1. Structural model of the pristine BiOI(001) slab: (a) side view showing the layered structure and the internal I···I van der Waals gap; (b) top view showing the surface atomic arrangement.

# 3. Results and Discussion

## 3.1. Adsorption Characteristics: Interaction Strength vs Electronic Perturbation

To elucidate the microscopic origin of the gas sensing behavior of layered BiOI, density functional theory (DFT) calculations were performed to systematically investigate the adsorption of $NO_2$, $NH_3$, $CO_2$, $CH_3OH$, $CH_3COCH_3$, and $C_2H_5OH$ on the BiOI (001) surface. The (001) facet was selected due to its thermodynamic stability and its prevalence in experimentally synthesized layered BiOI nanostructures, making it highly relevant for practical sensing applications. The oxygen terminated BiOI (001) surface employed in this work preserves the intrinsic layered structural characteristics of BiOI, including the Bi–O and iodide layers, while providing accessible adsorption sites for gas molecules during adsorption and transport calculations.[51–53]. For each gas molecule, the most energetically favorable adsorption configuration was identified (Figure 2), and the corresponding adsorption energy ($E_{ads}$), charge transfer (ΔQ), interfacial bond length, and band gap variation ($\Delta E_g$) were evaluated, as summarized in Table 1. The bond-length labels in Table 1 identify the interacting atoms separately: for example, N($NO_2$)–O(surface) = 1.31 Å denotes an intermolecular gas–surface separation, not the N–O bond within $NO_2$. The calculated adsorption energies reveal a pronounced variation in gas surface interaction strength. $NH_3$ exhibits the strongest adsorption (−1.25 eV), followed by $NO_2$ (−0.91 eV), $CH_3COCH_3$ (−0.89 eV), and $CH_3OH$ (−0.78 eV). $C_2H_5OH$ shows comparatively moderate interaction (−0.53 eV), whereas $CO_2$ displays weak adsorption energy (−0.05 eV), indicative of physisorption. $NH_3$, $NO_2$, and $CH_3COCH_3$ exhibit comparatively stronger interaction with the BiOI (001) surface, with adsorption energies ranging from −0.89 to −1.25 eV. However, their adsorption characteristics differ significantly in terms of bonding configuration and interfacial geometry. $NH_3$ and $NO_2$ preferentially interact with surface oxygen atoms through nitrogen atoms, forming short N–O bonds of 1.41 Å and 1.30 Å, respectively. These short bond distances indicate strong local orbital overlap and pronounced surface interaction. In contrast, $CH_3COCH_3$ adsorbs through oxygen coordination with surface Bi atoms, forming a longer Bi–O bond of 2.56 Å. Part of this bond elongation can be attributed to the substantially larger atomic radius of Bi compared with N, since Bi possesses a much larger covalent radius. Nevertheless, the increased interfacial distance still suggests comparatively weaker orbital hybridization and reduced localized surface perturbation relative to $NH_3$ and $NO_2$ adsorption. This indicates that adsorption behavior cannot be interpreted solely from adsorption energy but should also consider bonding configuration and structural relaxation. Importantly, this classification highlights that adsorption strength alone does not fully determine sensing performance. Although $NH_3$ exhibits the strongest binding energy, the resulting electronic perturbation and its influence on charge transport depend not only on interaction strength but also on the nature of orbital hybridization and the modification of delocalized surface states. Therefore, understanding gas sensing behavior in layered BiOI requires adsorption energetics to be analyzed together with charge transfer and transport modulation, as discussed in the following sections.

Table 1. Calculated adsorption energies ($E_{ads}$ in eV), charge transfer (Δ Q in e) nearest gas–surface atomic distances ($d$, Å) and changes in the band gap ($\Delta E_g$ in eV) for the most stable adsorption configurations of gas molecules on the BiOI (001) surface. The distance labels explicitly identify the interacting gas and surface atoms; for example, N($NO_2$)–O(surface) denotes the distance between nitrogen in $NO_2$ and oxygen in BiOI, rather than an intramolecular N–O bond.

| Molecules | $E_{ads}$ (eV) | Δ Q (e) | Nearest gas–surface distance ($d$, Å) | $\Delta E_g$(eV) |
|---|---|---|---|---|
| $NO_2$ | −0.91 | 0.03 | N–O, 1.30 | 0.1 |
| $NH_3$ | −1.25 | −0.01 | N–O, 1.41 | 0.3 |
| $CO_2$ | −0.05 | −0.02 | C–O, 1.33 | 0.0 |
| $CH_3OH$ | −0.78 | 0.13 | Bi–O, 2.44 | 0.1 |
| $CH_3COCH_3$ | −0.89 | −0.26 | Bi–O, 2.56 | 0.1 |
| $C_2H_5OH$ | −0.53 | −0.28 | Bi–O, 3.17 | 0.2 |

Under O-rich chemical-potential conditions, the O-containing O/I surface exhibits a surface excess energy of 0.00774 eV Å$^{-2}$, below that of the reconstructed I/I surface (0.11032 eV Å$^{-2}$). The O/I termination is therefore favored by 0.10258 eV Å$^{-2}$ within this O-rich limit and provides the surface model used for the adsorption and transport calculations.

The internal interlayer spacing was obtained as the average perpendicular separation between the two facing iodine planes across the I···I van der Waals gap in each relaxed BiOI(001) slab (Table 2). The pristine value of 4.119 Å decreases to 3.848 Å for $NO_2$, 3.796 Å for $NH_3$, 3.776 Å for $CO_2$, 3.815 Å for $CH_3OH$, 4.088 Å for $CH_3COCH_3$, and 3.712 Å for $C_2H_5OH$, corresponding to contractions of 6.6%, 7.8%, 8.3%, 7.4%, 0.8%, and 9.9%, respectively. The contraction does not vary monotonically with adsorption energy or the calculated current response. Thus, interlayer relaxation is a possible contribution to adsorption-induced changes in the electronic structure and transport, but the present results do not establish it as the sole or dominant mechanism.

Table 2. Interlayer van der Waals gap ($d_{vdW}$) of pristine and gas-adsorbed BiOI(001), defined as the average perpendicular separation between the two facing iodine planes. The absolute change (Δd) and percentage change are calculated relative to pristine BiOI. Negative values indicate interlayer contraction.

| System | I···I gap (Å) | Δd (Å) | Change (%) |
|---|---|---|---|
| Pristine BiOI | 4.119 | 0.000 | 0.0 |
| $NO_2$/BiOI | 3.848 | −0.271 | −6.6 |
| $NH_3$/BiOI | 3.796 | −0.323 | −7.8 |
| $CO_2$/BiOI | 3.776 | −0.343 | −8.3 |
| $CH_3OH$/BiOI | 3.815 | −0.304 | −7.4 |
| $CH_3COCH_3$/BiOI | 4.088 | −0.031 | −0.8 |
| $C_2H_5OH$/BiOI | 3.712 | −0.407 | −9.9 |

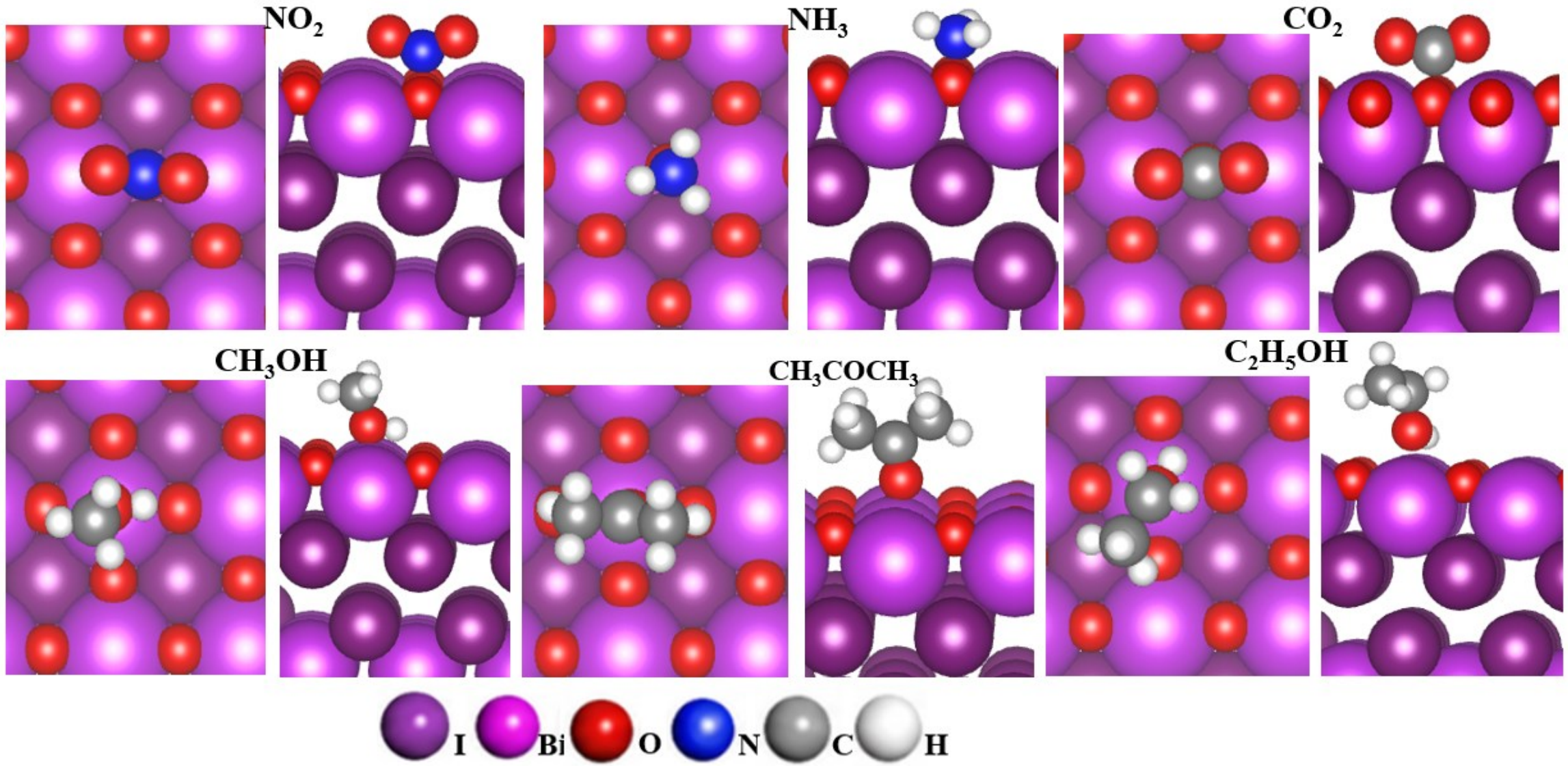

Figure 2. Top and side views of the most stable adsorption configurations of $NO_2$, $NH_3$, $CO_2$, $CH_3OH$, $CH_3COCH_3$, and $C_2H_5OH$ molecules on the BiOI (001) surface.

### 3.1.2 Charge Transfer and Carrier Modulation

Charge transfer analysis provides critical insight into the sensing mechanism of BiOI by quantifying both the direction and magnitude of electron exchange between the adsorbed molecules and the surface[54]. Since BiOI exhibits p-type semiconducting behavior, variations in surface electron density directly influence hole concentration and, consequently, electrical conductivity. As summarized in Table 1, $NO_2$ exhibits a positive charge transfer ($\Delta Q = +0.03$ $|e|$), indicating electron withdrawal from the BiOI surface. As an electron acceptor, $NO_2$ increases the surface hole concentration in p-type BiOI. In contrast, $NH_3$ shows a negative charge transfer ($\Delta Q = -0.01$ $|e|$), corresponding to electron donation to the surface. Although $NH_3$ exhibits the strongest adsorption energy among the investigated gases, its net charge redistribution remains limited, suggesting that the interaction is dominated by localized bonding rather than extensive perturbation of delocalized surface states. Consequently, strong adsorption does not necessarily translate into significant carrier density modulation. $CO_2$ also exhibits charge transfer ($\Delta Q = -0.02$ $|e|$), despite displaying the weakest adsorption energy. This indicates that charge transfer magnitude alone is insufficient to explain sensing behavior, and that transport modulation additionally depends on how adsorption affects transmission channels and electronic state delocalization near the Fermi level. Among the volatile organic compounds, more distinct charge transfer behavior is observed. $CH_3OH$ acts as an electron acceptor with a relatively large positive charge transfer ($\Delta Q = +0.13$ $|e|$), whereas $CH_3COCH_3$ and $C_2H_5OH$ donate electrons to the surface, exhibiting substantial negative charge transfer values of −0.26 |e| and −0.28 |e|, respectively. These variations reflect differences in molecular polarity, adsorption geometry, and bonding configuration, leading to gas dependent modulation of surface carrier concentration.

Importantly, the results demonstrate that adsorption strength and charge redistribution are not linearly correlated. For example, $NH_3$ exhibits the strongest binding energy but only minimal net charge transfer, whereas some moderately adsorbed VOCs induce comparatively larger electron exchange. This decoupling indicates that adsorption energy alone is insufficient to predict sensing performance. Overall, the magnitude and direction of charge transfer play a central role in determining conductivity trends in p-type BiOI. Electron accepting species enhance hole concentration, while electron donating molecules partially compensate surface holes. However, as will be shown in subsequent transport analysis, carrier modulation should be interpreted together with electronic structure perturbation and transmission channel evolution to fully understand sensing behavior.

## 3.2 Electronic Structure Perturbation

Gas adsorption induces modifications in the electronic structure of BiOI, which can be assessed through band gap variation and density of states (DOS) analysis (Figure 3). Among the investigated molecules, $NH_3$ produces the largest band-gap change (0.3 eV), followed by $C_2H_5OH$ (0.2 eV), while $NO_2$ and other VOCs induce moderate variations (~0.1 eV). In contrast, $CO_2$ adsorption results in negligible band gap change, consistently with its weak physisorption. These variations arise from adsorption induced surface states and charge redistribution near the Fermi level, both of which directly influence carrier transport in chemiresistive sensing. Figure 4 presents the total DOS of gas molecules on the BiOI (001) surface before and after gas adsorption. Noticeable gas dependent perturbations are observed near the Fermi level (0 eV), where transport relevant electronic states reside. For strongly adsorbed $NO_2$, pronounced modifications appear close to the Fermi level, indicating significant hybridization between molecular orbitals and surface states. Such hybridization introduces localized states that can act as carrier scattering centers, thereby modulating electrical transport. $NH_3$ also induces detectable DOS changes; however, the perturbation near the Fermi level is comparatively weaker, consistent with its limited net charge transfer despite strong binding.

In contrast, $CO_2$ adsorption produces only minor DOS variations and does not introduce significant mid gap states. The preservation of electronic states near the Fermi level suggests that the delocalized conduction channels of BiOI remain largely intact. VOC molecules exhibit intermediate behavior, with moderate DOS modifications reflecting partial orbital overlap but less pronounced electronic perturbation. Overall, the DOS analysis demonstrates that the impact of adsorption on transport properties depends not only on interaction strength but also on the extent to which electronic states near the Fermi level are perturbed. Strong hybridization can introduce localized scattering states, whereas weak physisorption preserves extended conduction pathways an effect that becomes crucial in understanding the bias dependent transport behavior discussed in the following section.

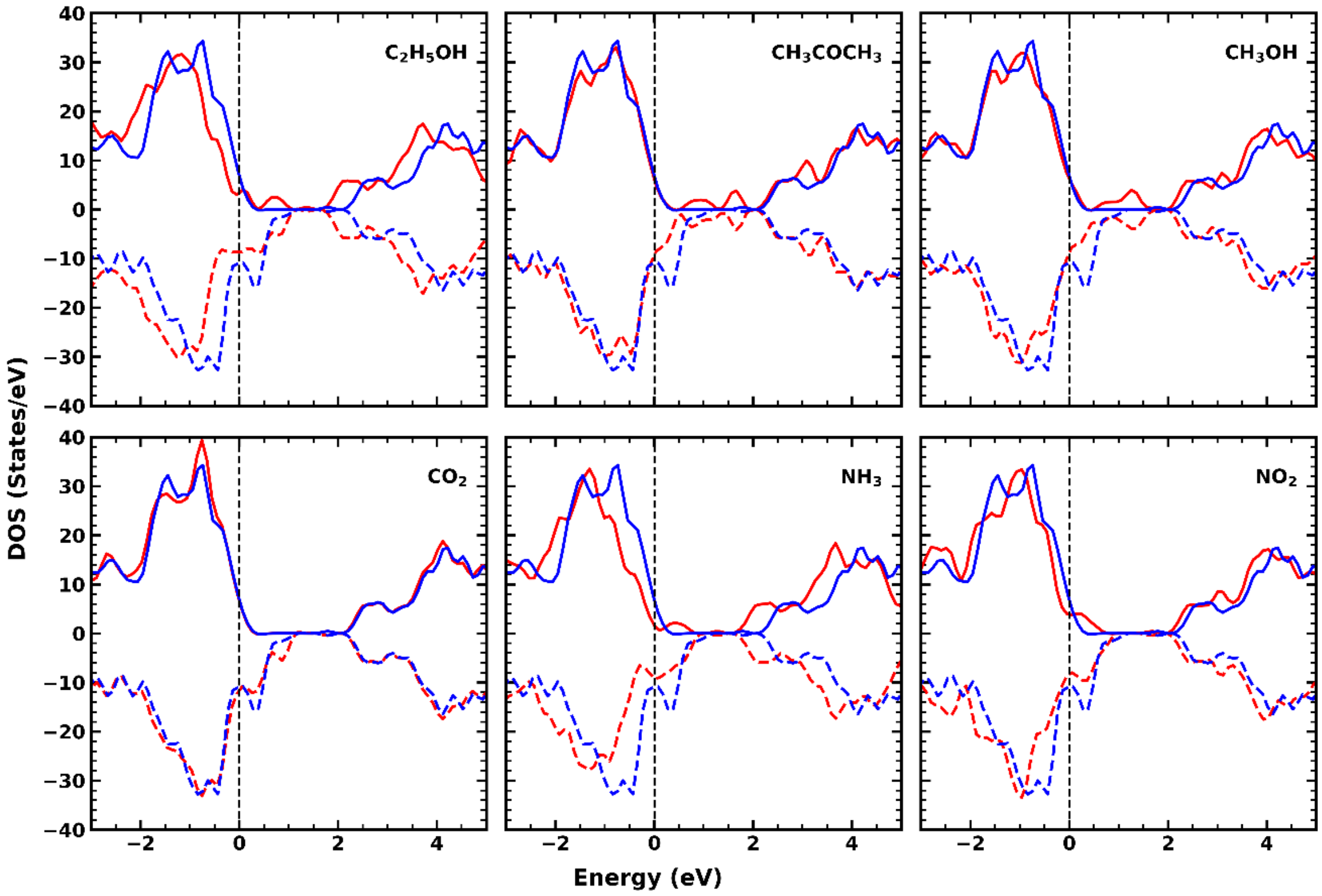

Figure 3. Spin-resolved density of states (DOS) of pristine BiOI(001) and BiOI(001) following adsorption of $NO_2$, $NH_3$, $CO_2$, $CH_3OH$, $CH_3COCH3$, and $C_2H_5OH$. Red and blue curves represent the gas-adsorbed and pristine surfaces, respectively. Solid and dashed lines denote the spin-up and spin-down channels, respectively. The Fermi level is set to 0 eV.

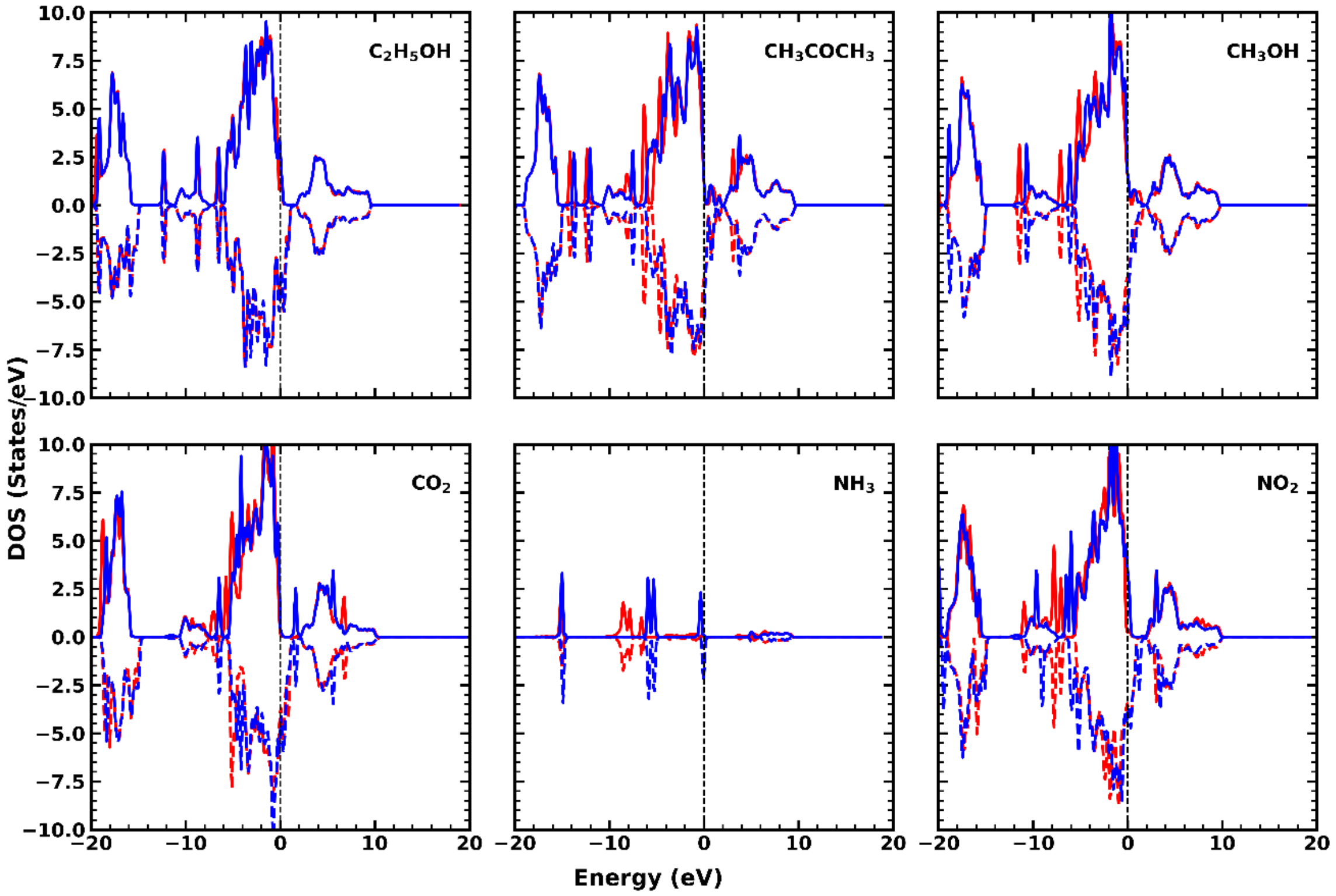

Figure 4. Spin-resolved density of states (DOS) of $NO_2$, $NH_3$, $CO_2$, $CH_3OH$, $CH_3COCH_3$, and $C_2H_5OH$ molecules before and after adsorption on BiOI(001). Red curves represent the DOS of the adsorbed molecules, while blue curves represent the DOS of the isolated molecules before adsorption. Solid and dashed lines denote the spin-up and spin-down channels, respectively. The Fermi level is set to 0 eV.

While the present data do not consider spin–orbit coupling, see Figure S2a for an example showing its minor effect on the calculated data (almost perfectly retained DOS profile for the occupied states below $E_F$ and to a large extent also for unoccupied states above $E_F$). The electronic band structure of pristine BiOI(001) along the Γ–X–M–Γ high-symmetry is presented in Figure 5. Note that the spin-polarized calculation yields a small but non-zero magnetic moment (predominantly associated with p orbitals of Bi) of ≈0.006 $\mu_B$ per atom (≈0.021 $\mu_B$ per Bi atom). This is probably due to symmetry breaking introduced by the surface termination and should not be confused with bulk characteristics of diamagnetic BiOI.

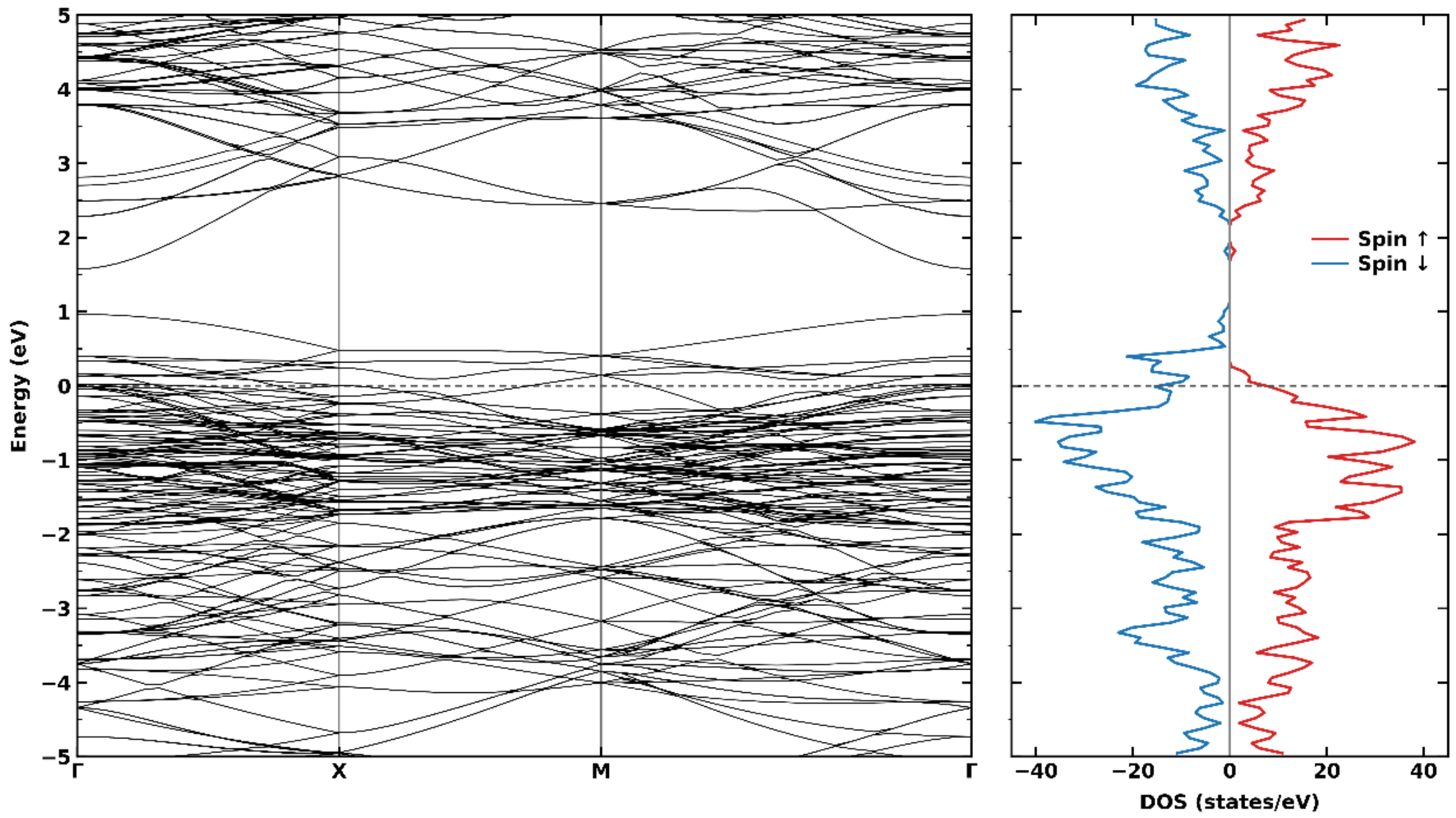

Figure 5. Calculated electronic band structure and spin-resolved density of states (DOS) of pristine BiOI(001). The band structure is shown along the Γ–X–M–Γ high-symmetry path, while the PBE-calculated DOS is shown on the right.

## 3.3 Bias Dependent Transport Modulation

To directly correlate adsorption behavior with sensing response, the charge transport properties of the BiOI (001) surface before and after gas adsorption were evaluated using the non-equilibrium Green's function (NEGF) formalism. The transport setup is illustrated in Figure 6 Transport properties are a critical indicator of gas sensing performance, as gas adsorption directly modifies carrier concentration, scattering mechanisms, and interfacial barrier heights in chemiresistive sensors[55]. The resulting current voltage (I–V) and resistance voltage (R–V) characteristics are presented in Figure 7a, b, respectively.

### 3.3.1 Pristine BiOI

The pristine BiOI surface exhibits an approximately linear I–V relationship within the applied voltage range (0.5–2 V), indicating quasi-ohmic transport without the formation of Schottky barriers. This linearity suggests efficient carrier injection between the electrodes and the scattering region, as well as structural stability of the transport pathway[56]. The resistance correspondingly decreases with increasing bias, which can be attributed to field assisted carrier transport and reduced contact resistance.

### 3.3.2 Gas Dependent Current Modulation

Although quasi-linear transport behavior is preserved after gas adsorption, distinct differences in current magnitude emerge depending on the adsorbed gas species. Among all investigated molecules, $CO_2$ adsorption produces the highest current across the applied bias range, whereas $NH_3$ results in the lowest current response. $NO_2$ and the VOC molecules exhibit intermediate behavior, while the pristine surface remains between the $CO_2$ and $NH_3$ adsorbed systems. These differences indicate that gas adsorption significantly modulates carrier transport without disrupting the intrinsic conductive pathway of the BiOI surface. At low bias (0.5 V), $CO_2$ exhibits the largest conductance modulation relative to the pristine surface despite its weak adsorption energy and minimal charge transfer (Figure 7). This behavior suggests that low bias sensing is governed primarily by transmission channel preservation rather than strong adsorption strength. As the applied bias increases, the relative transport responses of the adsorbed systems become more comparable, indicating that the sensing hierarchy is electrically tunable under different operating conditions. The resistance voltage characteristics further support this behavior. $NH_3$ adsorption yields the highest resistance values, whereas $CO_2$ maintains the lowest resistance. $NO_2$ and VOC-exposed systems exhibit intermediate resistance modulation. The clear separation of the I–V and R–V curves demonstrate that BiOI can effectively differentiate between oxidizing, reducing, and weakly interacting gases through distinct transport signatures.

### 3.3.3 Mechanistic Interpretation

The apparent paradox of weak adsorption but strong low bias sensitivity for $CO_2$ can be understood by analyzing transmission behavior near the Fermi level. $CO_2$ adsorption largely preserves the delocalized electronic states and transmission channels of the BiOI surface, introducing minimal scattering. As a result, carrier injection remains efficient, and even small perturbations can produce measurable conductance modulation under low bias. In contrast, strongly chemisorbed species such as $NH_3$ and $NO_2$ introduce localized states near the Fermi level, as revealed by DOS analysis. These localized states act as carrier scattering centers, partially suppressing transmission and reducing current at low bias. While strong adsorption enhances local electronic coupling, it can simultaneously hinder extended conduction pathways. Therefore, sensitivity in layered BiOI is governed by bias dependent transmission modulation rather than adsorption energy alone. Weak physisorption may preserve conductive channels and yield strong low bias response, whereas strong chemisorption can reduce current through scattering induced suppression of transport. This transport centered mechanism represents the key insight of the present work and provides a framework for electrically tunable gas selectivity at ambient conditions. The bias range up to 2.0 V was intentionally considered to examine the evolution of the intrinsic transport response and to facilitate comparison with previous DFT–NEGF studies employing a similar voltage range[57]. In the Au–BiOI–Au junction, the applied bias is distributed across the electrodes, metal–BiOI interfaces, and the BiOI region. Within the coherent-transport framework adopted here, these calculations provide a meaningful assessment of bias-dependent changes in transmission and gas response. Possible thermal degradation and electrical breakdown are device-level effects beyond the scope of this study.

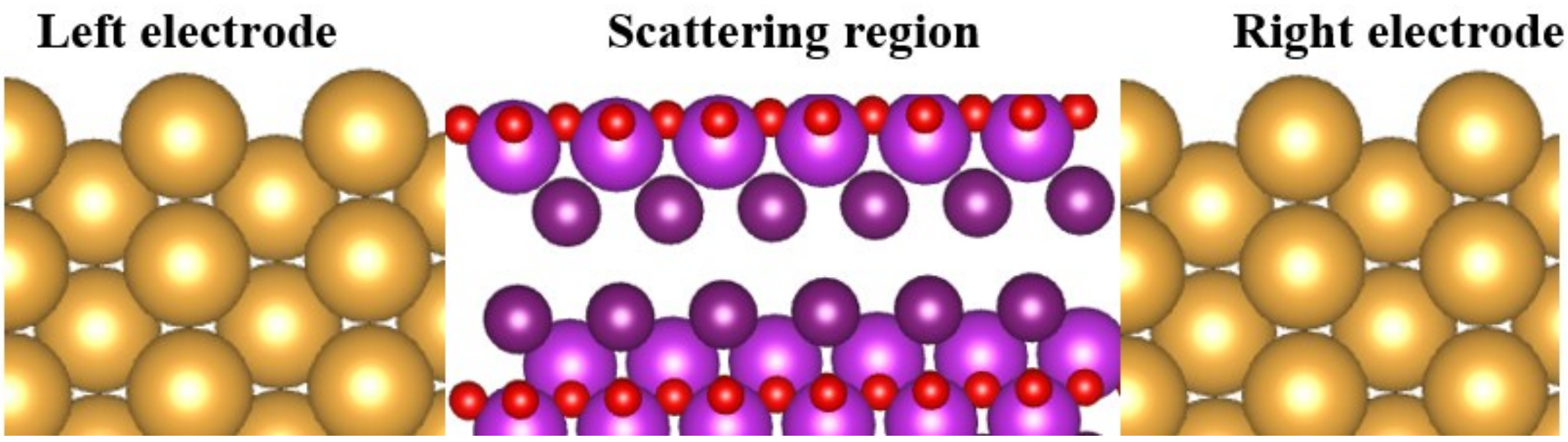

Figure 6. Schematic illustration of the two probe device configuration used for NEGF transport calculations. The BiOI (001) layer forming the scattering region is sandwiched between semi-infinite Au electrodes (left and right), which act as charge reservoirs for electron injection and collection. Gold electrodes are employed due to their high electrical conductivity, chemical stability, and ductility, ensuring reliable electrical contact and mechanical robustness of the device structure.

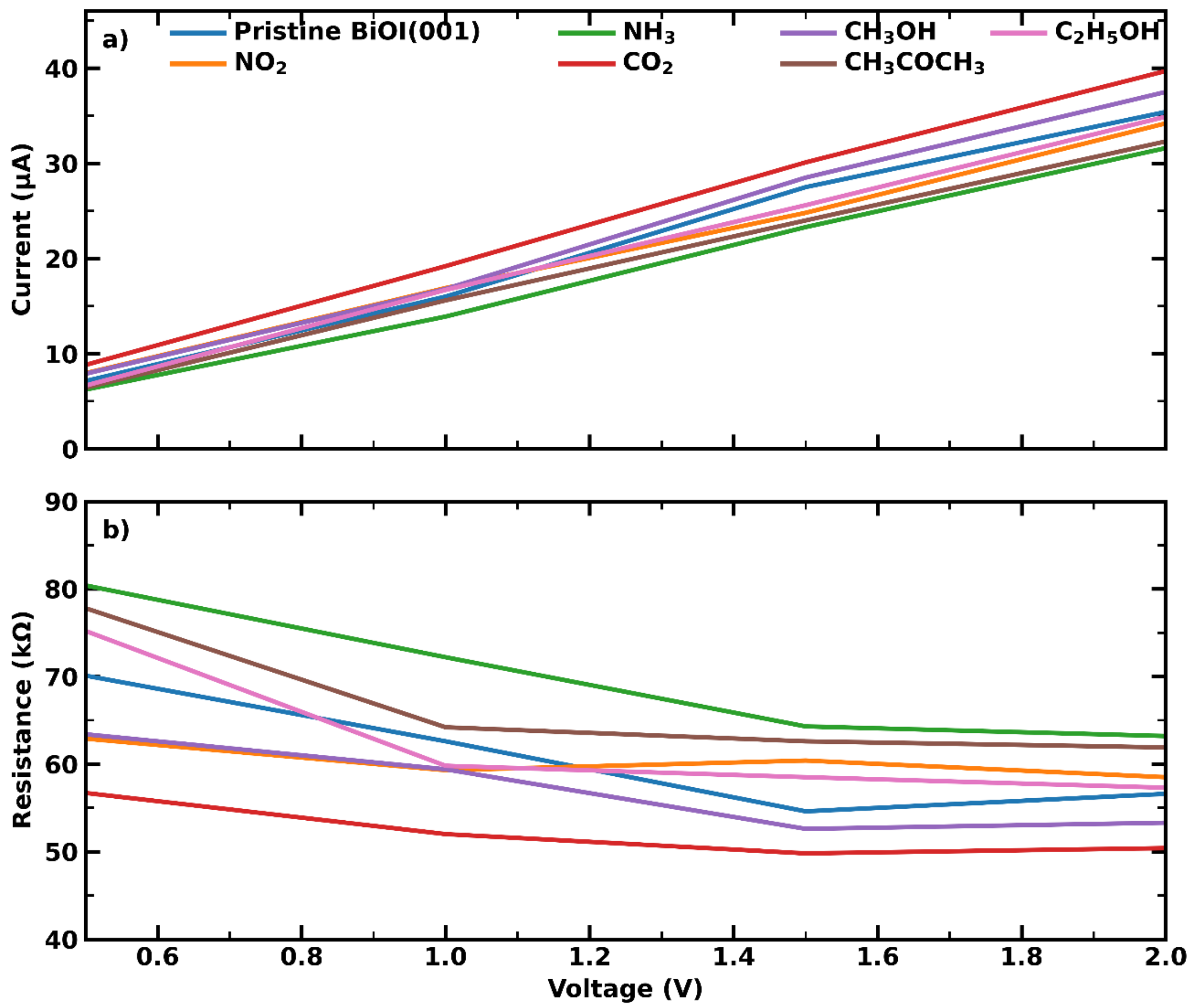

Figure 7. a) I-V and b) R-V characteristics of BiOI (001) device with and without adsorption of $NO_2$, $NH_3$, $CO_2$, $CH_3OH$, $CH_3COCH_3$, and $C_2H_5OH$.

### 3.4 Transmission Spectra Analysis

Zero-bias transmission spectra provide further insight into the microscopic origin of the observed transport behavior. Figure 8 compares the transmission functions of the pristine BiOI (001) device and those after adsorption of representative gas molecules.

The pristine surface exhibits finite transmission near the Fermi level ($E_F = 0$ eV), confirming the presence of available conductive channels for carrier transport. Upon $CO_2$ adsorption, the transmission intensity around the Fermi level remains comparable to or slightly higher than that of the pristine system. This preservation of transmission peaks indicates that the delocalized conduction pathways of BiOI remain largely intact. Because $CO_2$ interacts weakly with the surface and induces minimal charge redistribution, it does not introduce significant scattering centers, thereby maintaining efficient carrier injection and higher current in the I–V characteristics. In contrast, $NH_3$ adsorption noticeably suppresses the transmission intensity near the Fermi level. This reduction reflects the introduction of localized states and enhanced carrier scattering, consistent with the DOS analysis. As an electron donor, $NH_3$ partially compensates holes in the p-type BiOI surface, reducing the number of effective conducting channels and leading to lower current. Overall, the zero bias transmission analysis confirms that preservation of delocalized electronic states is beneficial for low bias sensing. Systems that maintain transmission channels near the Fermi level (e.g., $CO_2$) exhibit higher conductance, whereas adsorption induced scattering and localization (e.g., $NH_3$) suppress transport. These results reinforce that gas selectivity in BiOI is governed by transmission channel evolution rather than adsorption strength alone. Finite-bias transmission spectra of pristine BiOI(001), $CO_2$/BiOI(001), and $NH_3$/BiOI(001) at 0.0, 0.5, and 1.5 V are presented in Figure S1. The different bias-dependent transmission changes observed for different molecules provide further insight into the bias-dependent gas responses. Because current depends on transmission integrated over the bias window, these results complement the calculated I–V characteristics.

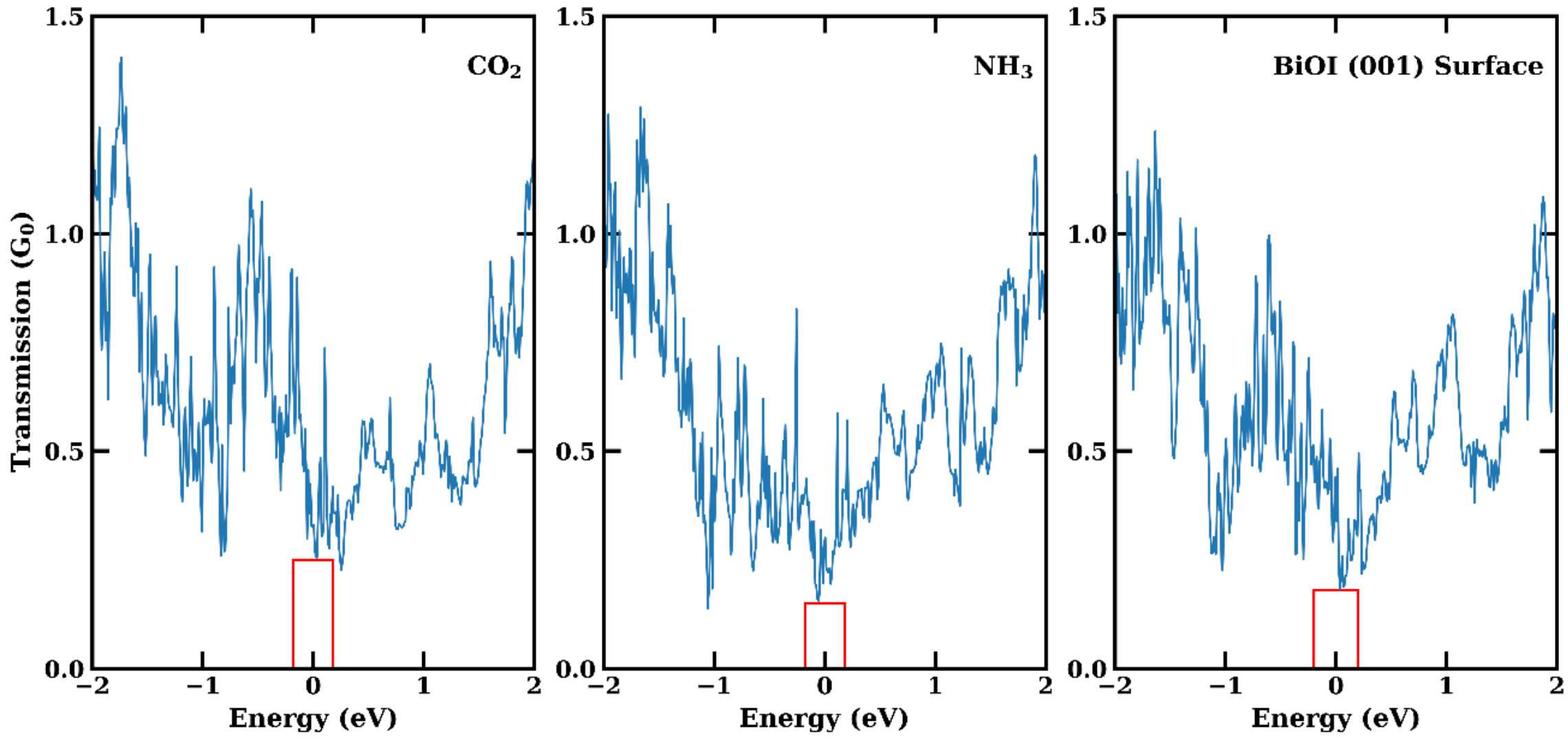

**Figure 8.** Zero-bias transmission spectra of the pristine BiOI (001) device and the systems adsorbed with $CO_2$ and $NH_3$ molecules. The red rectangular markers highlight the transmission region near the Fermi level ($E_F$ = 0), where adsorption induced modulation of conductive channels is most significant for charge transport behavior.

## 3.5 Sensitivity and Recovery Time Trade Off

The sensing performance of the BiOI (001) surface was quantified using the normalized conductance (or current) variation:

$$S = \frac{\mid G - G_0 \mid}{G_0} = \frac{\mid I - I_0 \mid}{I_0}$$

where $G_0$ and $I_0$ represent the conductance and current of the pristine BiOI surface, and $G$ and $I$ correspond to the values after gas adsorption. This definition is widely adopted in chemiresistive gas sensing studies[58] because it directly reflects the modulation of charge transport induced by gas surface interactions. That said, let us recall that the response at a particular surface concentration is not to be confused with a particular concentration in the atmosphere (the relationship between the former and the latter depends on the adsorption energies given in Table 1). The calculated sensitivities (Figure 9) exhibit strong dependence on both gas species and applied bias. At low voltage (0.5 V), the sensitivity hierarchy follows: $CO_2$>$NH_3$>$NO_2$>VOCs. $CO_2$ displays the highest sensitivity (0.23) despite its weak adsorption energy (which may limit its surface population even when an adsorbed molecule changes the device transmission), confirming that low bias sensing is governed primarily by transmission channel preservation rather than strong chemisorption. As the applied voltage increases, the hierarchy shifts. At 1.5 V: $NH_3$>$CH_3COCH_3$>$NO_2$≈$CO_2$. This bias dependent reordering demonstrates that gas selectivity in BiOI is electrically tunable. $NH_3$ shows enhanced sensitivity at intermediate bias due to field assisted modulation of hole transport, whereas $CO_2$ maintains moderate sensitivity across the entire voltage range, consistent with its minimal scattering and preserved conduction pathways. In contrast, $NO_2$ and VOCs exhibit intermediate or weaker responses depending on the balance between charge transfer and scattering effects.

Furthermore, note that $S = |I - I_0|/I_0$ denotes the magnitude of the calculated electrical response, following the aforementioned definition of $S$ in the literature[58]. The signed quantity $(I - I_0)/I_0$ additionally retains the direction of the current change (see Table S2 in Supporting information for signed data). While the $S$ values predict gas-dependent transport fingerprints and bias-dependent response differentiation for separately adsorbed gases, this should not be confused with selectivity measured in mixtures at controlled concentrations.

To evaluate reversibility, recovery times (τ) were estimated using an Arrhenius type desorption model[59]:

$$\tau = \nu^{-1} \exp\left(\frac{-E_{\mathrm{ads}}}{kT}\right)$$

where ν is the attempt frequency (typically ~$10^{12}$–$10^{13}$ $s^{-1}$), $E_{ads}$ is the adsorption energy, $k$ is the Boltzmann constant, and $T$ = 300 K. This expression highlights the exponential dependence of recovery time on adsorption energy, making τ extremely sensitive to gas surface interaction strength at room temperature. Strongly chemisorbed species such as $NH_3$ and $NO_2$ yield long theoretical recovery times at room temperature, indicating kinetically hindered desorption[51, 52]. In contrast, $CO_2$ exhibits ultrafast recovery consistent with its weak physisorption (Table 3). This disparity highlights a fundamental trade off in room temperature gas sensing. From a practical perspective, BiOI is intrinsically well suited for reversible detection of weakly interacting gases under ambient conditions. For strongly adsorbed species, external stimuli such as mild thermal activation or photo-assisted desorption may be required to accelerate recovery. Overall, the combined sensitivity and kinetic analysis demonstrates that optimal sensing performance in layered BiOI requires balancing adsorption strength with transmission channel preservation and desorption kinetics. This reinforces that transport modulation rather than adsorption energy alone governs effective gas selectivity. The recovery time was estimated using the Arrhenius-type relation $\tau = \nu^{-1} \exp(-E_{ads}/kT)$, with an assumed attempt frequency $\nu = 10^{12}$ $s^{-1}$, temperature $T$ = 300 K, and $E_{ads}$ approximated by the magnitude of the adsorption energy. The results are, of course, based on the same approximations as the Arrhenius formula itself (not considering, for example, the effects of gas concentration, surface coverage and competing adsorption).

Table 3. Recovery time (τ in second) of gas molecules on BiOI (001) surface at room temperature

| Molecules | $E_{ads}$ (eV) | $\exp(^{-E_{ads}}/_{kT})$ | τ (s) |
|---|---|---|---|
| $NO_2$ | −0.91 | $1.7 \times 10^{15}$ | $1.7 \times 10^{3}$ |
| $NH_3$ | −1.25 | $1.8 \times 10^{21}$ | $1.8 \times 10^{9}$ |
| $CO_2$ | −0.05 | 6.9 | $6.9 \times 10^{-12}$ |
| $CH_3OH$ | −0.78 | $1.1 \times 10^{13}$ | $1.1 \times 10^{1}$ |
| $CH_3COCH_3$ | −0.89 | $9.0 \times 10^{14}$ | $9.0 \times 10^{2}$ |
| $C_2H_5OH$ | −0.53 | $8.0 \times 10^{8}$ | $8.0 \times 10^{-4}$ |

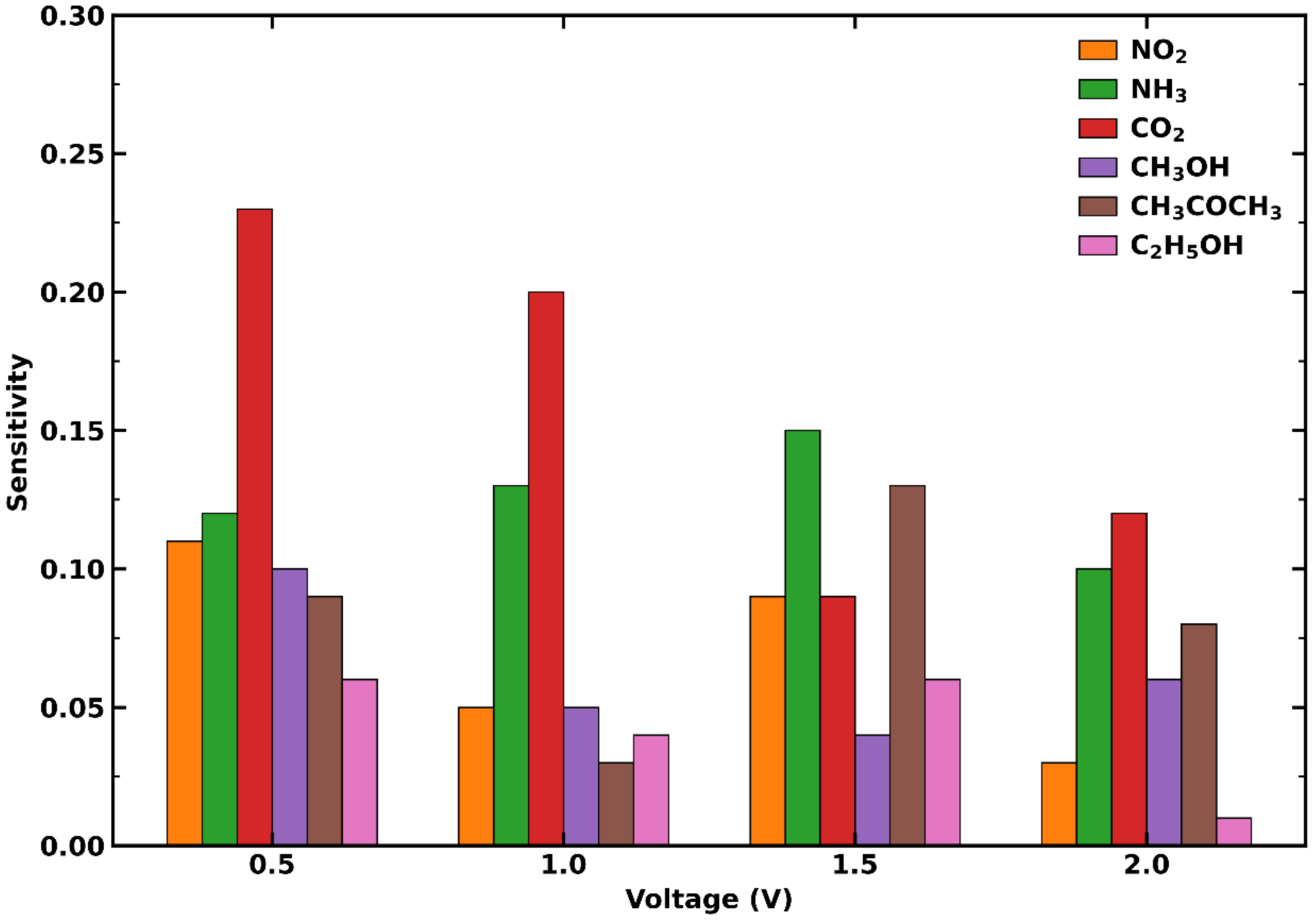

Figure 9. Bias-dependent sensitivity of $NO_2$, $NH_3$, $CO_2$, $CH_3OH$, $CH_3COCH_3$, and $C_2H_5OH$ adsorbed on the BiOI (001) surface calculated from current modulation under different applied voltages.

## 3.6 Transport Centered Selectivity Framework

The combined analysis of adsorption energetics, charge transfer, density of states, transmission spectra, and bias dependent sensitivity reveals a unified transport centered sensing mechanism in layered BiOI. Gas adsorption induces charge redistribution at the surface, the direction and magnitude of which depend on the molecular electronic structure and bonding configuration. This adsorption induced electronic perturbations modify the electronic states near the Fermi level and consequently influence carrier transport in the p-type BiOI surface. More importantly, adsorption alters the nature of transport channels near the Fermi level. Depending on the interaction strength and adsorption geometry, transmission pathways can either remain largely preserved, maintaining delocalized conduction channels, or become partially suppressed through the formation of localized scattering states. These modifications directly affect carrier injection and transport under applied bias. The application of an external electric field further amplifies these differences. Because the transport response depends on both adsorptions induced electronic perturbation and bias driven carrier dynamics, the sensitivity hierarchy is not fixed but becomes electrically tunable. As demonstrated in the preceding sections, weakly adsorbed species may dominate at low bias due to preserved transmission channels, whereas more strongly interacting gases can induce enhanced transport modulation at intermediate bias. Therefore, gas selectivity in layered BiOI is fundamentally governed by bias dependent transport modulation rather than adsorption energy alone. This transport centered framework provides a comprehensive description of gas sensing behavior in layered semiconductors and offers a rational basis for designing electrically tunable gas sensing platforms. Experimental studies on mesoporous BiOI nanoflakes have reported comparatively strong ethanol sensing [35], while BiOI-containing heterostructures have demonstrated $NO_2$ detection[34]. Because the experimental devices and the present idealized Au–BiOI–Au model differ in surface morphology, concentrations, and contact conditions, direct numerical agreement in response magnitude or recovery time is not expected. To put an example of a testable transport prediction, the relative $CO_2$ response exceeds the $NH_3$ response at 0.5 V, with the ordering reversing near 1.5 V under the modeled adsorption configurations.

## 4. Conclusions

This work establishes a transport centered framework for understanding bias dependent gas sensing in layered BiOI through the combined use of density functional theory and non-equilibrium Green's function analysis. Although strong chemisorption is observed for $NH_3$ and $NO_2$, adsorption strength alone does not determine sensing performance. Instead, the sensing hierarchy is governed by bias dependent modulation of transmission channels and charge carrier scattering near the Fermi level. Weakly adsorbed $CO_2$ preserves delocalized conductive pathways and exhibits pronounced low bias sensitivity, whereas strongly bound species introduce localized states that partially suppress transmission and kinetically hinder recovery. The demonstrated ability to electrically tune sensitivity and gas selectivity underscores the importance of coupling adsorption energetics with transport analysis when evaluating gas sensing materials. Overall, these findings provide mechanistic insight into electric field controlled sensing in layered semiconductors and suggest that optimal sensor design requires balancing adsorption strength, transmission channel preservation, and desorption kinetics for low power and reversible gas detection.

## Declaration of Generative AI and AI-assisted technologies in the writing process

During the preparation of this work the authors used Grammarly and ChatGPT to improve the grammar and style of the draft text. After using this service, the authors reviewed and edited the content as needed and take full responsibility for the publication's content.

## Acknowledgement

The DFT calculation was supported by the project QM4ST, funded as project No. CZ.02.01.01/00/22_008/0004572 by Programme Johannes Amos Comenius, call Excellent Research. Additionally, this work was supported by the project COLOSSE funded by the European Union under Grant Agreement No. 101158464.

Table S1. Convergence tests for bulk BiOI: (a) variation of the electronic total energy with k-point mesh and (b) variation of the electronic total energy and optimized lattice parameters (a, b, and c) with plane-wave energy cutoff (ENCUT).

(a)

| k-point mesh | Electronic total energy (eV) |
|---|---|
| 4 x 4 x 4 | -26.50 |
| 5 x 5 x 5 | -26.46 |
| 6 x 6 x 6 | -26.47 |
| 7 x 7 x 7 | -26.47 |
| 8 x 8 x 8 | -26.47 |

(b)

| ENCUT (eV) | Electronic total energy (eV) | a (Å) | b (Å) | c (Å) |
|---|---|---|---|---|
| 350 | -26.90 | 3.88 | 3.88 | 8.77 |
| 400 | -26.55 | 3.96 | 3.96 | 8.98 |
| 450 | -26.65 | 3.99 | 3.99 | 9.12 |
| 500 | -26.66 | 4.00 | 4.00 | 9.27 |
| 550 | -26.68 | 4.01 | 4.01 | 9.39 |
| 600 | -26.70 | 4.02 | 4.02 | 9.58 |

Table S2. Calculated sensitivity of $NO_2$, $NH_3$, $CO_2$, $CH_3OH$, $CH_3COCH3$, and $C_2H_5OH$ on BiOI (001) surface.

| Species | Sensitivity | | | | |
|---|---|---|---|---|---|
| | 0 V | 0.5 V | 1.0 V | 1.5 V | 2.0 V |
| $NO_2$ | 0 | 0.11 | 0.05 | -0.09 | -0.03 |
| $NH_3$ | 0 | -0.12 | -0.13 | -0.15 | -0.10 |
| $CO_2$ | 0 | 0.23 | 0.20 | 0.09 | 0.12 |
| $CH_3OH$ | 0 | 0.10 | 0.05 | 0.04 | 0.06 |
| $CH_3COCH_3$ | 0 | -0.09 | -0.03 | -0.13 | -0.08 |
| $C_2H_5OH$ | 0 | -0.06 | 0.04 | -0.06 | -0.01 |

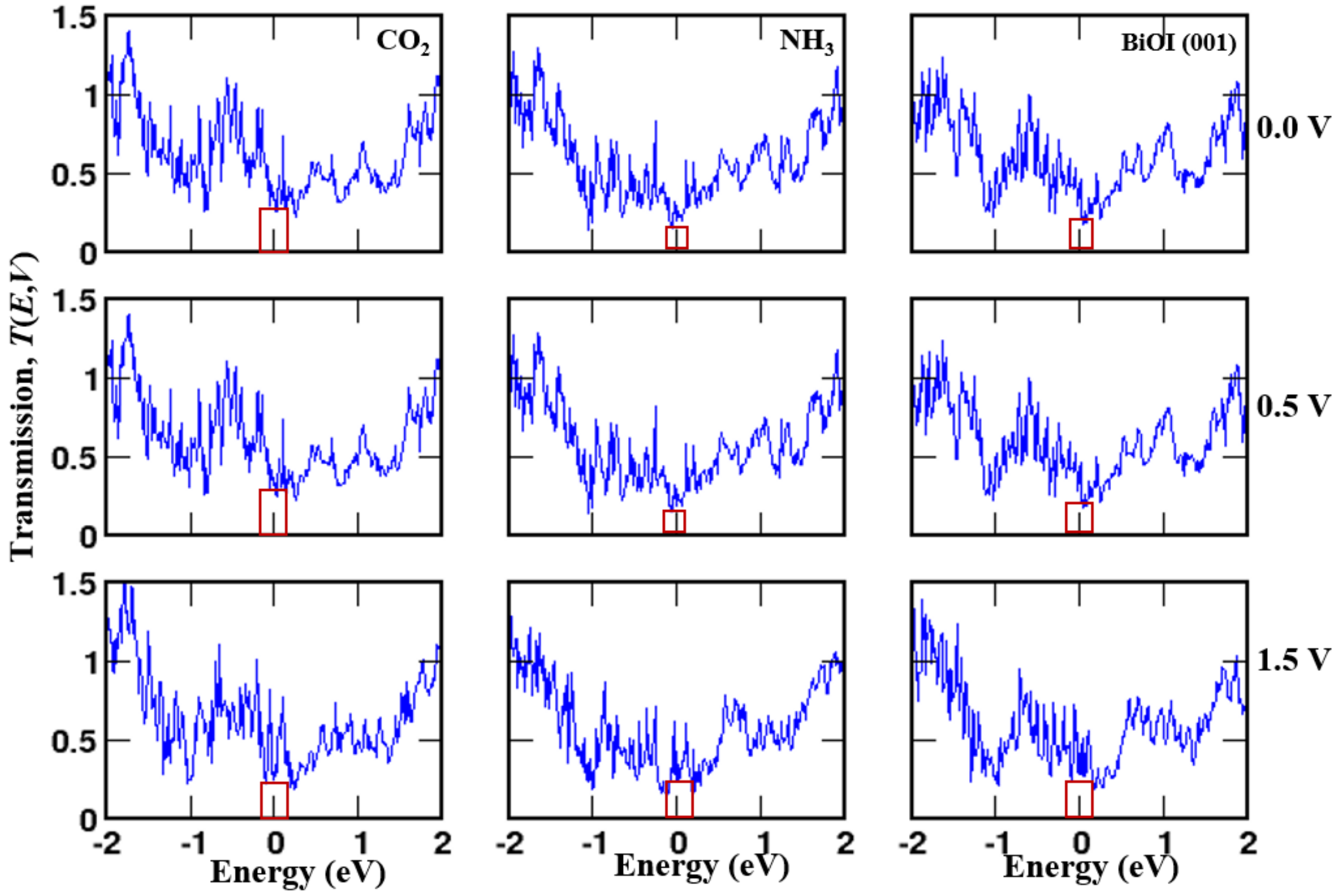

Figure S1. Electron transmission spectra, $T\ (E,\ V)$, of the Au–BiOI–Au device for $CO_2$-adsorbed BiOI(001) (left column), $NH_3$-adsorbed BiOI(001) (middle column), and pristine BiOI(001) (right column). The top, middle, and bottom rows correspond to source–drain biases of 0.0, 0.5, and 1.5 V, respectively.

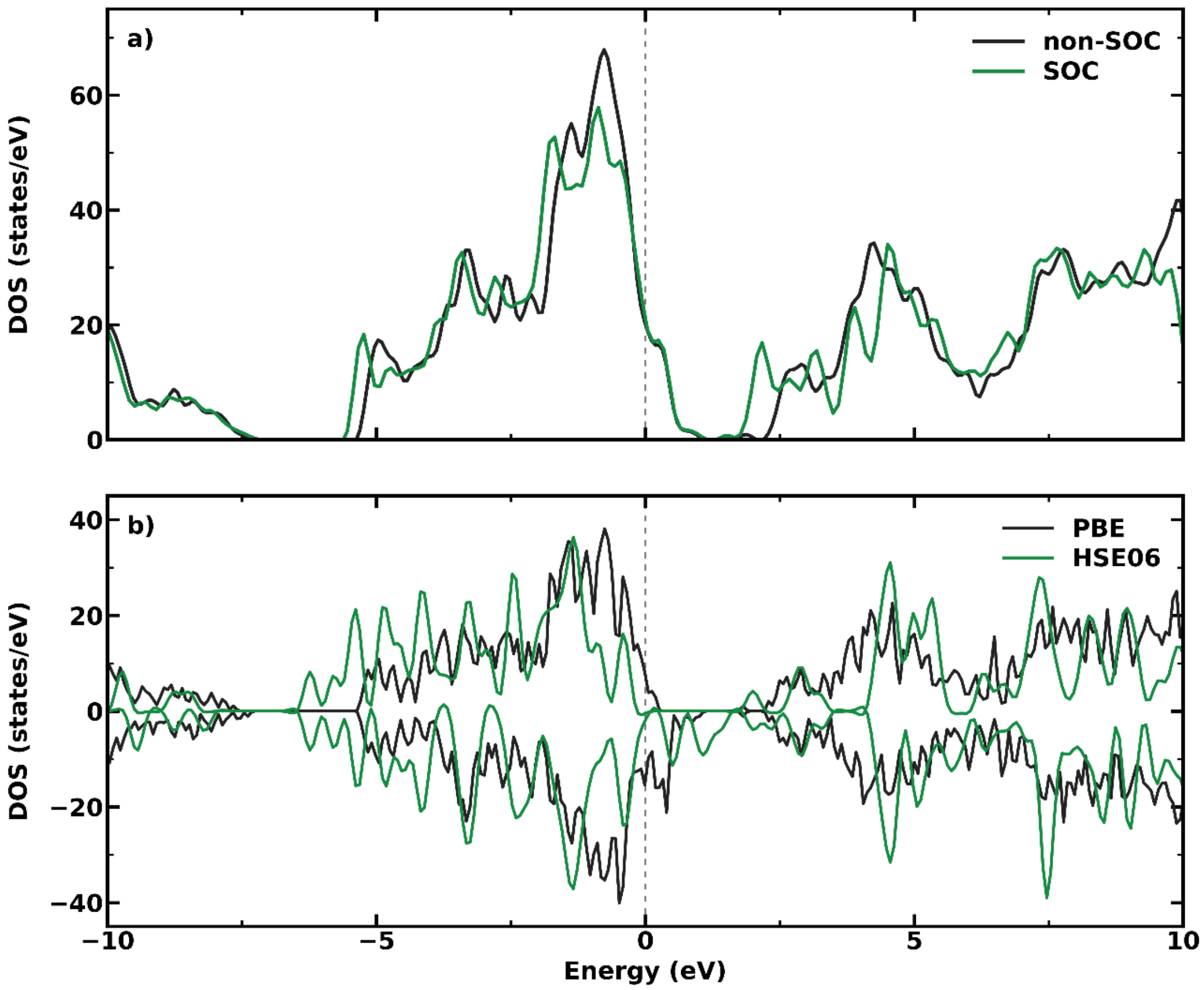

Figure S2. Calculated density of states (DOS) of pristine BiOI(001). (a) Comparison of the total DOS with and without spin–orbit coupling (SOC). (b) Spin-resolved DOS calculated using the PBE and HSE06 functionals.